\documentclass[aps,pra,onecolumn,superscriptaddress,12pt,notitlepage]{revtex4-1}
\usepackage{graphicx}
\usepackage{amsmath} 
\usepackage{amssymb} 
\usepackage{bm}
\usepackage[dvips]{color}
\usepackage{xspace}
\usepackage{latexsym} 
\usepackage[mathscr]{eucal} 
\usepackage{color}
\usepackage[paperwidth=210mm,paperheight=297mm,centering,hmargin=2.5cm,vmargin=2.5cm]{geometry}

\newcommand{\tr}[1]{%
{}^t\hspace{-0.8mm}#1%
}

\newcommand{\Tr}{\mathrm{Tr}}
\newcommand{\lv}{\left \vert}
\newcommand{\rv}{\right \vert}
\newcommand{\la}{\left \langle}
\newcommand{\ra}{\right \rangle}
\newcommand{\ket}[1]{\lv #1 \ra}
\newcommand{\bra}[1]{\la #1 \rv}

\newcommand{\ie}{\textit{i.e.}\xspace}

\newcommand{\I}{\mathbb{I}}
\newcommand{\R}[1]{\vec{R}_{#1}}

\newcommand{\Hilb}[1]{\mathcal{H}_{#1}}
\newcommand{\LOCC}{\mathrm{LOCC}}

\newtheorem{Definition}{Definition}
\newtheorem{Theorem}{Theorem}

\newtheorem{Lemma}{Lemma}

\newtheorem{Corollary}{Corollary}

\begin{document}
\title{Comparing globalness of bipartite unitary operations: delocalization power, entanglement cost, and entangling power}
\author{Akihito Soeda}
\affiliation{Department of Physics, Graduate School of Science,
The University of Tokyo, Tokyo 113-0033, Japan.}
\author{Mio Murao}
\affiliation{Department of Physics, Graduate School of Science,
The University of Tokyo, Tokyo 113-0033, Japan.}
\affiliation{Institute for Nano Quantum Information Electronics,
The University of Tokyo, Tokyo 113-0033, Japan}

\begin{abstract}
We compare three different characterizations of the globalness of bipartite unitary operations, namely,  delocalization power, entanglement cost, and entangling power, to investigate global properties of unitary operations. We show that the globalness of the same unitary operation depends on whether input states are given by unknown states representing pieces of quantum information, or a set of known states for the characterization.  We extend our analysis on the delocalization power in two ways.  First, we show that the delocalization power differs whether the global operation is applied on one piece or two pieces of quantum information.  Second, by introducing a new task called LOCC one-piece relocation,  we prove that the controlled-unitary operations do not have the delocalization power strong enough to relocate one of two pieces of quantum information by adding LOCC.
\end{abstract}

\maketitle

\section{Introduction}

Understanding the source of quantum advantage in quantum computation is a long-standing issue in quantum information science.  Previous researches have shown that certain quantum computation is `classical', for the reason that it is efficiently simulateable by classical computers.  One example is any computation performed just by local operations and classical communication (LOCC) \cite{Horodecki, Plenio} without using any entangled resources.  All models of quantum computation outperforming classical counterparts use entanglement resources (such as measurement-based quantum computation~\cite{Raussendorf}) or some kind of non-LOCC operation.  Non-LOCC operations are called `global' operations.  The source of quantum speedup must be due to the properties of the global operations.  In this paper, we refer to the properties exclusive to global operations as \textit{globalness} of quantum operations.

It is also known that not all global operations result in quantum speedup for quantum computation.  There must be a specific globalness that differentiates the quantum operations leading to quantum speedup from those do not.  The difference may be due to more than one kind of globalness, but even this is not clear at this point.  For this reason, having a good understanding of the globalness of quantum operations is important.  In this paper, we try to understand the simplest case of the global operations, namely, bipartite unitary operations.  

To investigate globalness of unitary operations, it is important to clarify what kind of states is given as inputs of the unitary operations.   We want to evaluate the globalness that does not depend on a choice of a particular input state.  By introducing the concept of {\it pieces of quantum information}, we analyze characterizations of unitary operations for two pieces of quantum information represented by arbitrary unknown states, in terms of {\it delocalization power} \cite{Soeda} and {\it entanglement cost} \cite{SoedaLOCC}. We compare these characterizations with another characterization, {\it entangling power} of global operations \cite{Kraus, Wolf, Linden}, which characterizes the globalness of unitary operations acting on a set of known states.

Then we extend our analysis of globalness in terms of the delocalization power in two ways by introducing new LOCC tasks.  One of the tasks is {\it LOCC one-piece relocalization} for {\it one piece} of delocalized quantum information that corresponds to the case when a part of input state is unknown and arbitrary but the other part can be chosen from a set of known state.  The other task is {\it LOCC one-piece relocation} for two pieces of delocalized quantum information, which evaluates the ability of the unitary operation to relocate one of the two pieces of quantum information from one Hilbert space to another by adding LOCC to the unitary operation. 

The rest of the paper is organized as following.  In Section \ref{overview}, we introduce the concept of pieces of quantum information and present an overview on the three characterizations.  We summarize the comparison of different aspects of the globalness of bipartite unitary operations presented in the previous works in Section \ref{comparison}.  We extend the analysis of the delocalization power in Sections \ref{fixed_input} and \ref{relocation}.  In Section \ref{fixed_input}, we show the result on LOCC one-piece relocalization for one piece of delocalized quantum information.  In Section \ref{relocation}, we analyze  LOCC one-piece relocation of two pieces of quantum information. Finally, in Section \ref{conclusion}, we present our conclusion.

\section{Three characterizations of the globalness of quantum operations} \label{overview}

\subsection{Delocalization power} \label{delocalizationpwer}

First, we define {\it a piece of quantum information} for a $d$-dimensional quantum system, or \textit{qudit}, whose Hilbert space is denoted by $\Hilb{}= \mathbb{C}^d$. 
\begin{Definition}
If a pure quantum state of $n$ qudits $\ket{\psi} \in \Hilb{}^{\otimes n}$ is given by 
$$
 \ket{\psi} = \sum_i \alpha_i \ket{\varphi_i},
$$
where $\{ \ket{\varphi_i} \}_{i=0}^{d-1}$ is a fixed set of normalized and mutually orthogonal states in $\Hilb{}^{\otimes n}$ and the coefficients $\alpha_i \in \mathbb{C}$ are arbitrary and unknown except for the normalization $\sum_i |\alpha_i|^2 = 1$, the unknown state $\ket{\psi}$ is said to represent {\it one piece} of quantum information for a qudit.  
\end{Definition}

In the formalism presented above, a piece of quantum information for a single qudit can be stored in an $n$-qudit system using an arbitrary set of orthonormal states, $\{ \ket{\varphi_i} \}_{i=0}^{d-1}$.  Any such set of states would form a \textit{logical} qudit space, but in a special case satisfying
\[
 \ket{\varphi_i} = \ket{i} \otimes \ket{\xi}
\]
for all $i \in \{ 0,\ldots,d-1\}$, where the set of states $\{  \ket{i} \}$ forms an orthonormal basis of $\Hilb{}$ and $\ket{\xi} \in \Hilb{}^{\otimes (n-1)}$ is independent of $i$, the piece of quantum information is stored in a \textit{physical} qudit.    Hence it is possible to {\it assign} one physical qudit for each piece of quantum information.  

Using this formalism, now we provide the formal definition of one piece of {\it localized} quantum information for a qudit.  We label the qudits of an $n$-qudit system from $0$ to $n-1$ and denote the Hilbert space of qudit $k$ by $\Hilb{k}$.  The Hilbert space of $n-1$ qudits \textit{excluding} a certain qudit $k$ will be denoted by $\Hilb{\lnot k}$.  We will also assume that two different pieces of quantum information in the same system are assigned to different physical qudits.
\begin{Definition}   
For  $n \geq 2$, a piece of quantum information represented by an unknown $n$-qudit state $\ket{\psi}$ is said to be {\it localized} at {\it an assigned} Hilbert space $\Hilb{k}$, or simply localized when there is no fear of confusion,  if it is represented in the form
$$
\ket{\psi} = \sum_i \alpha_i \ket{i} \otimes \ket{\xi},
$$
where $\{ \ket{i} \}_{i=0}^{d-1}$ is any basis of the Hilbert space of the assigned qudit (\ie, $\Hilb{k}$), $\ket{\xi} \in \Hilb{\lnot k}$ is an $(n-1)$-qudit state determined independently of the set of coefficients $\{ \alpha_i \}$, and $\{ \alpha_i \}$ are arbitrary coefficients satisfying the normalization condition $\sum_i |\alpha_i|^2 = 1$.  \end{Definition}

Note that the global phase factor of the coefficients is not a physical quantity, so we take the global phase equivalence.   There are $d-1$ complex degrees of freedom in total for one piece of quantum information.  For $n=1$, since $\Hilb{1} = \mathbb{C}^d$ is the minimal Hilbert space to store one piece of quantum information for a qudit, one piece of quantum information has to be localized in $\Hilb{1}$.

We define the concept of {\it delocalized} quantum information, which is the complement of localized quantum information, and also the concept of {\it delocalization} of quantum information.
\begin{Definition}
If a piece of quantum information is \textit{not} localized, then it is said to be \textit{delocalized}.  The task of delocalizing quantum information is called \textit{delocalization}.
\end{Definition}

Next, we consider two-qudit states, where the state of each qudit represents one piece of localized quantum information.  We denote the two Hilbert spaces of the qudits by $\Hilb{A}= \mathbb{C}^d$ and $\Hilb{B}= \mathbb{C}^d$. The two pieces of localized quantum information can be represented by a tensor product state $\ket{\psi_A}_A \otimes \ket{\psi_B}_B$, where the subscripts of the kets denote the assignment of the Hilbert spaces of the qudits, $\ket{\psi_A}_A \in \Hilb{A}$ and $\ket{\psi_B}_B \in \Hilb{B}$. We define delocalization for two pieces of quantum information as the following.
\begin{Definition}
Two pieces of quantum information are said to be \textit{delocalized}, if the state representing the two pieces of quantum information {\it cannot} be written by $u_A \ket{\psi_A}_A \otimes u_B \ket{\psi_B}_B$, where $u_A$ on $\Hilb{A}$ and $u_B$ on $\Hilb{B}$ are arbitrary local unitary operations but independent of $\ket{\psi_A}_A$ and  $\ket{\psi_B}_B$. 
\end{Definition}
We again note that we have already assigned a Hilbert space for each piece of quantum information, so the state $\ket{\psi_B}_A \otimes \ket{\psi_A}_B$ represents delocalized two pieces of quantum information out from the assigned Hilbert spaces. 

Now we investigate the effects of a global unitary operation $U$ applied on two pieces of localized quantum information  $\ket{\psi_A}_A \otimes \ket{\psi_B}_B$.  If the unitary operation $U$ is not a tensor product of two local unitary operations on $\Hilb{A}$ and $\Hilb{B}$, $U$ always transforms each piece of localized quantum information to delocalized quantum information.  In this paper, we say that the unitary operations have {\it delocalization power}, which in a sense is the `strength' of delocalization of quantum information due to the unitary operations.

How pieces of quantum information are delocalized is determined only by the set of orthonormal states representing the quantum information, which in turn is determined by the unitary operation used for the delocalization.  Therefore, the globalness of unitary operations can be studied by understanding how a unitary operation delocalizes pieces of quantum information.

Later, we argue that certain pieces of quantum information are `more' delocalized than others.  The difference in the level of delocalization can only have come from the difference in the globalness of the unitary operation, namely, the delocalization power.  Hence, we can classify the delocalization power by analyzing the level of the delocalization that each unitary operation achieves.
   
To define and classify the level of delocalization, we introduce the following LOCC task, {\it LOCC one-piece relocalization}, that aims to localize just one of the two pieces of delocalized quantum information by sacrificing the other piece of quantum information in $\Hilb{A} \otimes \Hilb{B}$.  We denote the set of density operators on the Hilbert space  $\Hilb{}$ by $\mathcal{S} ( \Hilb{})$.
\begin{Definition}
{\it LOCC one-piece relocalization} of qudit $B$ for two pieces of quantum information delocalized by a global unitary operation $U$ is a task to find an LOCC-implementable completely positive trace preserving (CPTP) map ${\Lambda}_U^{\LOCC}: \mathcal{S} ( \Hilb{A} \otimes \Hilb{B} ) \rightarrow \mathcal{S} (\Hilb{B} )$ satisfying
$$
\Lambda_U^{\LOCC} [U  (\ket{\psi_A}_A\bra{\psi_A} \otimes \ket{\psi_B}_B \bra{\psi_B})  U^\dag] = \ket{\psi_B}_B \bra{\psi_B}
$$
for any $\ket{\psi_A}_A \in \Hilb{A}$ and $\ket{\psi_B}_B \in \Hilb{B}$. 
\end{Definition}

We characterize the delocalization power of global unitary operations in terms of their ability to allow LOCC one-piece relocalization on two pieces of quantum information delocalized by the global unitary operations.   We define the order of the delocalization power of two global unitary operations $U$ and $U'$ on two pieces of quantum information by the following.
\begin{Definition}
If LOCC one-piece relocalization of two pieces of delocalized quantum information is possible for a unitary operation $U$, but not possible for another unitary operation $U'$, the order of the delocalization power of $U$ is defined to be smaller than that of $U'$ in terms of LOCC one-piece relocalization. 
\end{Definition}

\subsection{Entanglement cost}

Another way to quantify the globalness of a unitary operation applied on quantum information is to evaluate how much extra global resource is required on top of LOCC operations to implement the unitary operation on two pieces of quantum information. The minimum amount of entanglement required to deterministically implement a given global operation is unique, based on the fact that entanglement cannot be generated by LOCC.  We define an LOCC task, {\it entanglement assisted deterministic LOCC implementation} and then define {\it entanglement cost} of the unitary operation on quantum information in terms of this LOCC task.

\begin{Definition}
{\it Entanglement assisted deterministic LOCC implementation} of a global unitary operation $U$ on two pieces of localized quantum information $\ket{\psi_A}_A \otimes \ket{\psi_B}_B \in \Hilb{}^{\rm in} = \Hilb{A}^{\rm in} \otimes \Hilb{B}^{\rm in}$ using a fixed bipartite resource state $\ket{\Phi}_{AB} \in \Hilb{}^{\rm r} = \Hilb{A}^{\rm r} \otimes \Hilb{B}^{\rm r}$ is a task of finding an LOCC-implementable CPTP map $\Gamma_U^{\LOCC}: \mathcal{S} ( \Hilb{}^{\rm in}  \otimes \Hilb{}^{\rm r} ) \rightarrow \mathcal{S} ( \Hilb{}^{\rm in}  )$ satisfying
$$
\Gamma_U^{\LOCC} ( \ket{\psi_A}_A\bra{\psi_A} \otimes \ket{\psi_B}_B \bra{\psi_B}  \otimes \ket{\Phi}_{AB} \bra{\Phi}) = U (\ket{\psi_A}_A\bra{\psi_A} \otimes \ket{\psi_B}_B \bra{\psi_B}) U^\dag
$$
for any $\ket{\psi}_A \in \Hilb{A}^{\rm r} $ and $\ket{\psi}_B \in \Hilb{B}^{\rm r} $.
\end{Definition}

\begin{Definition}
{\it Entanglement cost} of a unitary operation $U$ on two pieces of quantum information is given by the minimum amount of entanglement of the resource state $\ket{\Phi}_{AB}$ required to perform entanglement assisted deterministic LOCC implementation of $U$ on two pieces of localized quantum information.
\end{Definition}

The entanglement cost of unitary operations can be regarded as the minimum entanglement cost for delocalizing two pieces of quantum information. Thus this is another way to characterize the globalness of unitary operations applied on quantum information. 

\subsection{Entangling power}

We can also quantify the globalness of a unitary operation by evaluating its ability of entanglement generation in place of entanglement cost, similarly to the pair of the definitions for evaluating the globalness of quantum states, namely, distillable entanglement and entanglement cost.  However, the amount of entanglement generated by a unitary operation strongly depends on the choice of input states, therefore it is difficult to define a quantity in terms of quantum information, namely, unknown states. Instead, {\it entangling power} of a global operation \cite{Kraus, Wolf, Linden} is defined by optimizing the generated amount of entanglement over a set of known input states.

\begin{Definition}
The entangling power of a bipartite unitary operation $U$ (denoted by $E_\mathrm{ep}(U)$) is defined as the maximum amount of entanglement generated between the bipartite cut by applying $U$ on a known state, \ie,
\[
 E_{\mathrm{ep}} (U) \equiv \max_{\rho_{\mathrm{in}}} E(U \rho_{\mathrm{in}} U^\dag) - E(\rho_{\mathrm{in}}),
\]
where $E(\rho)$ is an entanglement measure of choice and $\rho_{\mathrm{in}}$ is chosen among the given set of states.
 
\end{Definition}

\section{Comparison of globalness by different characterizations} \label{comparison}

In the previous section, we introduced three different characterizations for the globalness of bipartite unitary operations: delocalization power, entanglement cost and entangling power.  In this section, we summarize known results on the three characterizations and investigate whether the globalness characterized by each method is same to or different from that of by the others.  

\begin{Theorem} \label{theorem:qq1pr}
LOCC one-piece relocalization for two pieces of quantum information of qudits delocalized by a unitary operation $U$ is possible if and only if $U$ is a locally unitary equivalent to a controlled-unitary operation $C_{\{u_k \}}=\sum_k \ket{k}\bra{k} \otimes u_k$, where $\{ \ket{k} \}$ forms an orthonormal basis for one of the subsystems and $\{ u_k \}$ is a set of unitary operations on the other subsystem. \cite{Soeda}
\end{Theorem}

The characterization of globalness based on the delocalization power reveals that there are two classes of globalness for bipartite unitary operations, one class is a local unitary equivalent of a controlled-unitary operations, and the other class is all the rest of global unitary operations.  

\begin{Theorem}
For any given two-qubit controlled-unitary operation, its entanglement cost for entanglement assisted deterministic LOCC implementation is 1 ebit when the Schmidt number, the number of non-zero Schmidt coefficients, of the resource state is 2. \cite{SoedaLOCC}
\end{Theorem}

For other two-qubit unitary operations, the entanglement cost of the swap operation $U_{\rm SWAP}$, of which action is given by  $U_{\rm SWAP} \ket{\psi_A}_A \otimes \ket{\psi_B}_B = \ket{\psi_B}_A \otimes \ket{\psi_A}_B $ for any $\ket{\psi_A}$ and $\ket{\psi_B}$, is easily shown to be 2 ebit by considering the situation where the two input qubits are parts of maximally entangled states.  However, for more general operations, it is not easy to evaluate the minimum entanglement cost of entanglement assisted LOCC implementation, therefore, it is still unknown. 

The formulation of entangling power depends on the set of allowed input states and the measure of entanglement.  Entangling power is usually difficult to calculate because it involves two optimizations.  One is the maximization over all possible input states (usually taken to be separable or product states).  The other is the calculation of the amount of generated entanglement according to the chosen entanglement measure.  Even when the quantum operation is restricted to bipartite unitary operations, the exact value is obtained for only limited cases~\cite{Kraus, Chefles}.  For example, it is known that the CNOT operation $C_X$ ($C_{\{ u_k \}}$ where $u_0=\mathbb{I}$ and $u_1$ is given by the Pauli matrix $X$) has the entangling power of 1 ebit and the swap operation $U_{\rm SWAP}$ has the entangling power of 2 ebit when we allow to use ancilla qubits. 

Nevertheless, we can make a relatively generic statement about entangling power if the entanglement measure is continuous.  The statement is as follows.  The identity operation clearly generates no entanglement at all, hence its entangling power should be zero.  Invoking a continuity argument, there should be a set of operations in the neighborhood of the identity operation such that their entangling power is arbitrarily small.

However, when we evaluate the globalness in terms of the delocalization power and entanglement cost,  a fundamental difference arrises.   By using these two characterizations, all two-qubit controlled-unitary operations of the form 
\begin{equation} \label{C-u}
C_u = \ket{0}\bra{0} \otimes \mathbb{I} + \ket{1}\bra{1} \otimes u,
\end{equation}
where $u$ is a single qubit unitary operation, belong to the same class of globalness irrelevant to their entangling power.  Thus, two-qubit controlled-unitary operations and their local unitary equivalent operations belong to a distinct class from the the class of identity operation, even when the local unitary operations are close to the identity ($u \approx \mathbb{I}$), in contrast to the characterization in terms of entangling power.  

On the other hand, for general \textit{known}  bipartite pure qudit states, the LOCC convertibility condition between two states is known.
\begin{Theorem} \label{majorizationtheorem}
A bipartite state $\ket{\Psi}_{AB}$ can be transformed to another state $\ket{\Psi'}_{AB}$ by using only LOCC, if and only if the Schmidt coefficients of the $\ket{\Psi}_{AB}$ is majorized by those of $\ket{\Psi'}_{AB}$  \cite{Majorization}. 
\end{Theorem}

Since the LOCC conversion protocol depends on the choice of states $\ket{\Psi}_{AB}$ and $\ket{\Psi'}_{AB}$ (when LOCC conversion from $\ket{\Psi}_{AB}$ to $\ket{\Psi'}_{AB}$ is possible), it is essential that these states are known.   By taking $\ket{\Psi}_{AB} = \ket{\psi_A}_A \otimes \ket{\psi_B}_{B}$ and $\ket{\Psi'}_{AB} =  U \ket{\Psi}_{AB}$ and  using Theorem \ref{majorizationtheorem}, we can see that if we have a resource state of which Schmidt coefficients are equal to the those of $\ket{\Psi'}_{AB}$, it is possible to obtain $U \ket{\Psi}_{AB}$ by LOCC.   For two-qubit unitary operations, which can only create an entangled state with Schmidt number 2, the majorization condition is equivalent to the comparison of the amount of entanglement.
Thus, we have the following corollary.
\begin{Corollary}
The entanglement cost of the resource state for entanglement assisted deterministic LOCC implementation of unitary operations on a {\it given known} state $\ket{\psi_A}_A \otimes \ket{\psi_B}_B$ is given by the the amount of entanglement of $U \ket{\psi_A}_A \otimes \ket{\psi_B}_B$.
\end{Corollary}

This corollary gives a justification to define the entanglement cost for an entanglement assisted deterministic LOCC implementation on a set of known states by finding the largest minimum entanglement cost to perform $U$ over the set of input states.   In this case, the entanglement cost for a set of known states coincides to the entangling power of $U$.  

The results in this section indicate that there are several aspects of globalness in quantum operations.
It is particularly important to clarify the types of input states, known states or unknown states representing pieces of quantum information, for analyzing globalness of unitary operations, since they lead to a fundamental difference. 

\section{Delocalization power for one piece of quantum information} \label{fixed_input}

In the classification of globalness of unitary operations in terms of the delocalization power presented in the previous sections, we analyzed global properties of two pieces of delocalized quantum information.  That is, we analyzed the globalness of unitary operations totally independent of input states.   On the other hand, in Section \ref{delocalizationpwer}, we also defined {\it one piece} of delocalized quantum information.   This situation corresponds to the case where one of the two input qudits is in an arbitrary and unknown state, but that of the other qudit is in a {\it known} state, and we can choose the most suitable state for performing tasks.   In this section, we extend our analysis on globalness of unitary operations in terms of delocalization power to the case for one piece of quantum information.    

We define the task of LOCC one-piece relocalization of {\it one piece} of delocalized quantum information.
\begin{Definition}
{\it LOCC one-piece relocalization} of the qudit $B$ for {\it one piece} of quantum information delocalized by a global unitary operation $U$ is a task to find an LOCC-implementable CPTP map ${\Lambda}_U^{\LOCC}: \mathcal{S} ( \Hilb{A} \otimes \Hilb{B} ) \rightarrow \mathcal{S} (\Hilb{B} )$ and a state $\ket{\xi_A} \in \Hilb{A}$ satisfying
$$
 {\Lambda}_U^{\LOCC} [U  (\ket{\xi_A}_A\bra{\xi_A} \otimes \ket{\psi_B}_B\bra{\psi_B})  U^\dag] = \ket{\psi_B}_B \bra{\psi_B}
$$
for any $\ket{\psi_B}_B \in \Hilb{B}$.
\end{Definition}

We show the following lemma.
\begin{Lemma}
The global unitary operations that allow LOCC one-piece relocalization for one piece of delocalized quantum information is in a strictly wider class of global unitary operations than that allows LOCC one-piece relocalization for two pieces of delocalized quantum information.
\end{Lemma}

To prove this lemma, we present an example of two-qubit unitary operations, $U_{\rm ex}$, where LOCC one-piece relocalization is impossible for delocalized two pieces of quantum information,  but it becomes possible if one of the qubits is promised to be in a particular pure state.  
Let us take the computational basis, which is an orthonormal basis of the composite Hilbert space $\Hilb{A} \otimes \Hilb{B}$ given by $\{ \ket{i}_A \otimes \ket{j}_B \}_{i,j}$, where $\{\ket{i}_A \}$ and $\{\ket{j}_B \}$ are orthonormal base for $\Hilb{A}$ and $\Hilb{B}$, respectively.  The matrix representation of $U_{\mathrm{ex}}$ in the computational basis is given by 
\begin{equation*}
 U_{\mathrm{ex}} =   \left(%
  \begin{array}{c c c c}%
      1 & 0 & 0 & 0 \\%
      0 & 0 & 1 & 0 \\%
      0 & 1 & 0 & 0 \\%
      0 & 0 & 0 & -1%
    \end{array}%
  \right).%
\end{equation*}

First, we show that LOCC one-piece relocalization of the qubit $B$ for one piece of quantum information delocalized by $U_{\rm ex}$ is possible by presenting that $U_{\rm ex}$ can be simulated by a locally unitary equivalent operation to a controlled-unitary operation if qubit $A$ is set to a particular state.  We set the state $\ket{\xi_A}_A \in \Hilb{A}$ to be $\ket{+} =  \left ( \ket{0}+ \ket{1} \right) /\sqrt{2}$.  It is easy to check that 
for an arbitrary $\ket{\psi_B}_B$,
$$
 U_{\mathrm{ex}} \left( \ket{+}_A \otimes \ket{\psi_B}_B \right) = (H \otimes \I) \cdot C_X \left ( \ket{+}_A \otimes \ket{\psi_B}_B \right )
$$
where $C_X$ is a controlled-NOT operation and $H$ denotes the single-qubit Hadamard operation represented in the computational basis by
$$
 H = \frac{1}{\sqrt{2}} \left(
   \begin{array}{c c}
     1 & 1 \\
	 1 & -1
   \end{array}
 \right).
$$
(The same calculation can be done using the stabilizer formalism~\cite{Gottesman} by exploiting the fact that $U_{\mathrm{ex}}$ is a Clifford operation.)  Thus the action of $U_{\rm ex}$ can be simulated by $(H \otimes \I) \cdot C_X$, a locally unitary equivalent operation to the controlled-NOT operation, when we fix one of the qubits to be in the state $\ket{+}_A$.   

From Theorem \ref{theorem:qq1pr}, any operation which is locally unitary equivalent to a controlled-unitary operation is LOCC one-piece relocalizable for two pieces of delocalized quantum information.   Note that, if an LOCC protocol relocalizes two pieces of delocalized quantum information, the same protocol must also relocalize one piece of delocalized quantum information.  Therefore, $U_{\rm ex}$ is LOCC one-piece relocalizable for one piece of delocalized quantum information.

Next, we show that $U_{\mathrm{ex}}$ itself is not locally unitary equivalent to controlled-unitary operations, therefore it is {\it not} LOCC one-piece relocalizable for two pieces of delocalized quantum information.  To show this, we analyze the Cartan numbers for two-qubit unitary operations. 

It is known that any two-qubit unitary operator on $\Hilb{A} \otimes \Hilb{B}$ has the following Cartan decomposition \cite{Kraus},
$$
u_A \otimes u_B \cdot \exp[{i (\gamma_X X_{A} \otimes X_{B} + \gamma_Y Y_{A} \otimes Y_{B} +\gamma_Z Z_{A} \otimes Z_{B})}] \cdot v_A \otimes v_B,
$$
by taking appropriate local unitary operations $u_A$, $v_A$ on $\Hilb{A}$ and $u_B$, $v_B$ on $\Hilb{B}$, coefficients $0 \leq \gamma_k \leq \pi/4$ ($k=X,Y,Z$), where $X$, $Y$ and $Z$ denote the Pauli matrices and each subscript of the Pauli matrix indicates the corresponding Hilbert space.  In this decomposition, the nonlocal component of the unitary operation is represented by the set of coefficients $\{ \gamma_k \}$. In this paper, we refer $\gamma_k$ to be a Cartan coefficient, and the number of non-zero Cartan coefficients to be the Cartan number.  The Cartan number of a unitary operation cannot be changed by local unitary operations, and two unitary operations with different Cartan numbers cannot be locally unitary equivalent to each other~\cite{Nielsen}. 

The Cartan decomposition of $U_{\mathrm{ex}}$ is given by 
$$
U_{\mathrm{ex}} =  u_A \otimes u_B \cdot \exp [i \pi/4 (X_{A} \otimes X_{B} + Y_{A} \otimes Y_{B})] \cdot v_A \otimes v_B
$$
by using appropriate local unitary operators $u_A$, $v_A$, $u_B$ and $v_B$ \cite{Anders}.  Thus the Cartan number of $U_{\mathrm{ex}}$ is 2. On the other hand, the Cartan decomposition of a controlled phase operation $C_{S_\theta}=\ket{0}\bra{0} \otimes \I + \ket{1}\bra{1} \otimes S_\theta$ where the single-qubit phase operation $S_\theta$ is defined by $S_\theta = \ket{0}\bra{0}+{\rm e}^{i \theta} \ket{1}\bra{1}$ is given by 
$$
C_{S_\theta}=e^{-i {\theta}/{4}} S_{{\theta}/{2}} \otimes S_{{\theta}/{2}} \cdot \exp \left [i {\theta}/{4} (Z_{A} \otimes Z_{B}) \right ].
$$
Thus, the Cartan number of the controlled-phase operation is 1.   It is also known that any controlled-unitary operations $C_u$ is locally unitary equivalent to $C_{S_\theta}$, therefore, the Cartan number of the operations that are locally unitary equivalent to controlled-unitary operations is also 1.  Therefore, $U_{\mathrm{ex}}$ cannot be locally unitary equivalent to the controlled-unitary operations. 

Note that some unitary operations remain LOCC one-piece \textit{un}relocalizable even for one piece of delocalized quantum information.  Such an example is the swap operation $U_{\mathrm{SWAP}}$.  By performing $U_{\mathrm{SWAP}}$, even if one of the input qubit states is fixed to a particular known state, a piece of quantum information represented by the other qubit's unknown input state completely moves out from the original Hilbert space and stored in the other Hilbert space.  This phenomena  is an example of what we call a \textit{relocation} of a piece of quantum information. Once this relocation happens, it is not possible to relocalize the piece of quantum information back to the original Hilbert space by LOCC alone.  It requires 1 ebit of entanglement to relocalize the one piece of relocated quantum information on top of LOCC by using quantum teleportation \cite{Bennett}. 
 
\section{Relocation of quantum information} \label{relocation}

\subsection{LOCC one-piece relocation}
In the previous section, we briefly introduced the concept of \textit{{relocation}} of a piece of quantum information.   But actually, $U_{\mathrm{SWAP}}$ provides relocation of both two pieces of quantum information.   $U_{\mathrm{SWAP}}$ is the only unitary operation that has the delocalization power strong enough to relocate two pieces of quantum information simultaneously without any additional operation or resource.   A wider class of unitary operations, namely, the local unitary equivalents of $U_{\mathrm{SWAP}}$, also relocates two pieces of quantum information, if local operations are allowed as an extra operation.   To define and classify delocalization power of unitary operations in terms of relocation, we further relax the condition of the additional operations to LOCC and investigate LOCC relocatability of one of the two pieces of quantum information delocalized by unitary operations.

\begin{Definition}
{\it LOCC one-piece relocation}  from $\Hilb{A}$ to $\Hilb{B}$ for two pieces of quantum information delocalized by a unitary operation $U$ is the task to find an LOCC-implementable CPTP map $\Lambda^\LOCC_U:  \mathcal{S} ( \Hilb{A} \otimes \Hilb{B} ) \rightarrow \mathcal{S} (\Hilb{B} )$ satisfying
\begin{equation} \label{def:qq2pd}
\Lambda^\LOCC_U  [U (\ket{\psi_A}_A \bra{\psi_A} \otimes \ket{\psi_B}_B\bra{\psi_B})  U^\dag] = \ket{\psi_A}_B \bra{\psi_A}
\end{equation}
for any $\ket{\psi}_A \in \Hilb{A}$ and $\ket{\psi}_B \in \Hilb{B}$.
\end{Definition}

This is a task similar to LOCC one-piece relocalization for two pieces of quantum information, in the sense by sacrificing one of two pieces of quantum information, we obtain one piece of localized quantum information.   The difference between these tasks is the location of the piece of localized quantum information.   We define the order of the delocalization power of two global unitary operations $U$ and $U'$ on two pieces of quantum information in terms of LOCC one-piece relocation by the following.
\begin{Definition}
If LOCC one-piece relocation of two pieces of delocalized quantum information is possible for a unitary operation $U$, but not possible for another unitary operation $U'$,  the order of the delocalization power of $U$ is defined to be larger than that of $U'$ in terms of LOCC one-piece relocation. 
\end{Definition}
Note that for LOCC one-piece relocation, feasibility of the task implies more delocalization power, whereas for LOCC one-piece relocalization, feasibility of the task implies less delocalization power. 

As the first step to classify the delocalization power of global unitary operations in terms of LOCC one-piece relocation, we show that where the locally unitary equivalent class of controlled-unitary operations lies in this classification. 
\begin{Lemma}
If two pieces of quantum information are delocalized by an operation locally unitary equivalent to controlled-unitary operations, LOCC one-piece relocation is not possible. 
\end{Lemma}

To prove this lemma, we employ the formulation of LOCC using accumulated operators \cite{Soeda, SoedaLOCC}.  In the following subsections, we first summarize the formulation, and then show the proof by contradiction. 

\subsection{Formulation of LOCC using accumulated operators}

We adopt the standard formulation of LOCC~\cite{Donald}.  In a two-party scenario, Alice and Bob perform one local measurement operation in turns while exchanging the outcome of each measurement operation by classical communication.  The measurement operation at a particular turn is chosen according to all the outcomes by the both parties up to that turn, where the choice is made following a protocol agreed beforehand by the parties.  Strictly speaking, we may consider LOCC protocols which cannot be expressed in this form.  These protocols, however, can always be substituted by the protocols in this standard form.

Each local quantum operation can be described as a generalized measurement, which is represented by a set of operators $\{ M^{(r)} \}$ satisfying the completeness relation
$
 \sum_r M^{(r)\dag} M^{(r)} = \I.
$
There exists one operator for each outcome in the measurement, which is denoted by the superscript $r$.

We add a subscript to the outcome index, for example $r_k$, to specify to which measurement operation the index belongs.  In this notation, $r_k$ belongs to the $k$-th measurement operation in the sequence.  We use $\R{k} = (r_1,r_2,\ldots,r_k)$ to denote the set of measurement outcomes of the first $k$ measurement operations in the sequence.  The $(k+1)$-th measurement operation is a function of $\R{k}$ and we denote the set of operators describing this measurement operation by
$$
 \{M^{(r_{k+1}|\R{k})}\}_{r_{k+1}}.
$$

Let us denote Alice's measurement operations by $M_A^{(r_n|\R{n-1})}$ and Bob's by $M_B^{(r_n|\R{n-1})}$.  We set
$
 M_A^{(r_1|\R{0})} = M_A^{(r_1)}
$
and
$
 M_B^{(r_1|\R{0})} = M_B^{(r_1)}.
$
Note that $M_A^{(r_n|\R{n-1})}$ is an operator on $\Hilb{A}$ and $M_B^{(r_n|\R{n-1})}$ is on $\Hilb{B}$.  When $n$-th turn is Alice's turn then $(n+1)$-th turn is Bob's turn, which implies that Alice does not perform any operation during this $(n+1)$-th turn.  In this case, we set Alice's measurement operation to the identity operation, \ie,
$
 \{ M_A^{(r_{n+1}|\R{n})} \} = \{ \I \}.
$
If this $(n+1)$-th turn happens to be Bob's, then his measurement operation is set to the identity operation.

The effect of the measurement operations accumulates as an LOCC protocol proceeds.  The accumulated effect up to a particular turn is expressed by the product of all the measurement operators corresponding to all the measurement outcomes obtained up to that point.  Given a particular sequence of measurement outcomes $\R{n}$, we represent the accumulated effect corresponding to this sequence by an \textit{accumulated operator} $A^{\R{n}}$ defined by
$$
 A^{\R{n}} = \prod_{k=1}^n M_A^{(r_k|\R{k-1})}.
$$
Bob's accumulated operator will be denoted by $B^{\R{n}}$ defined by a similar way of $A^{\R{n}}$.

\subsection{Impossibility of relocation}

Let us focus on two-qubit controlled-unitary operations for simplicity.  The following argument can be extended to arbitrary two-\textit{qudit} controlled-unitary operations.   We prove by contradiction that LOCC one-piece relocation of two pieces of quantum information delocalized by any controlled-unitary operation is impossible.  Now, consider the following scenario where Alice has an extra ancilla qubit, whose Hilbert space is denoted by $\Hilb{a}$.  Let $\ket{\Phi}_{Aa}$ denote a maximally entangled state between Alice's input qubit and the ancilla qubit defined by
$$
 \ket{\Phi}_{Aa} = \frac{1}{\sqrt{2}} (\ket{0}_A \otimes \ket{0}_a + \ket{1}_A \otimes \ket{1}_a) \in \Hilb{A} \otimes \Hilb{a}.
$$

Suppose that there is an LOCC one-piece relocation protocol for the given controlled-unitary operation $C_u$ defined by Eq.\ (\ref{C-u}).  Let Alice set her input qubit and the ancilla in the state of $\ket{\Phi}_{Aa}$ while Bob's input remains arbitrary.  Alice and Bob perform $C_u$ and the LOCC protocol to complete the relocation of one piece of quantum information from $\Hilb{A}$ to $\Hilb{B}$.  Then Alice's ancilla qubit and Bob's input qubit are in the state of
$$
 \ket{\Phi}_{aB} = \frac{1}{\sqrt{2}}(\ket{0}_a \otimes \ket{0}_B + \ket{1}_a \otimes \ket{1}_B) \in \Hilb{a} \otimes \Hilb{B},
$$
which implies that
$$
\Lambda_{C_u}^\LOCC (C_u \ket{\Phi}_{Aa}\bra{\Phi} \otimes \ket{\psi_B}\bra{\psi_B} C_u^\dag) = \ket{\Phi}_{aB}\bra{\Phi}
$$
holds for an arbitrary $\ket{\psi_B} \in \Hilb{B}$.  Using the accumulated operator representation of $\Lambda_{C_u}^\LOCC$, we have
\begin{equation} \label{dislocated_ao}
 \Tr_A [\sum_{\R{n}} (A^{\R{n}} \otimes B^{\R{n}}) (C_u \ket{\Phi}_{Aa}\bra{\Phi} \otimes \ket{\psi_B}\bra{\psi_B} C_u^\dag) (A^{\R{n}} \otimes B^{\R{n}})^\dag] = \ket{\Phi}_{aB}\bra{\Phi}.
\end{equation}

We modify the LOCC protocol $\Lambda_{C_u}^\LOCC$ by adding an extra measurement operation by Alice described by
$ \{\ket{0}_A \bra{0}, \ket{0}_A \bra{1} \}$, just after Alice's final measurement.  We denote this modified protocol by ${\Lambda'}_{C_u}^\LOCC$.  Direct substitution reveals that
\begin{multline*} 
{\Lambda'}_{C_u}^\LOCC (C_u \ket{\Phi}_{Aa}\bra{\Phi} \otimes \ket{\psi_B}\bra{\psi_B} C_u^\dag)\\
 = \Tr_A [\sum_{\R{n}} (\ket{0}_A\bra{0}A^{\R{n}} \otimes B^{\R{n}}) (C_u \ket{\Phi}_{Aa}\bra{\Phi} \otimes \ket{\psi_B}\bra{\psi_B} C_u^\dag) (\ket{0}_A\bra{0}A^{\R{n}} \otimes B^{\R{n}})^\dag]\\
 + \Tr_A [\sum_{\R{n}} (\ket{0}_A\bra{1}A^{\R{n}} \otimes B^{\R{n}}) (C_u \ket{\Phi}_{Aa}\bra{\Phi} \otimes \ket{\psi_B}\bra{\psi_B} C_u^\dag) (\ket{0}_A\bra{1}A^{\R{n}} \otimes B^{\R{n}})^\dag]. 
\end{multline*}

Since the partial trace $\Tr_A$ is taken and the additional measurement introduced for the protocol ${\Lambda'}_{C_u}^\LOCC$ acts only on $\Hilb{A}$,  we have
$$
{\Lambda'}_{C_u}^\LOCC (C_u \ket{\Phi}_{Aa}\bra{\Phi} \otimes \ket{\psi_B}\bra{\psi_B} C_u^\dag) 
=
{\Lambda}_{C_u}^\LOCC (C_u \ket{\Phi}_{Aa}\bra{\Phi} \otimes \ket{\psi_B}\bra{\psi_B} C_u^\dag). 
$$ Thus we obtain 
$$
{\Lambda'}_{C_u}^\LOCC (C_u \ket{\Phi}_{Aa}\bra{\Phi} \otimes \ket{\psi_B}\bra{\psi_B} C_u^\dag) = \ket{\Phi}_{aB}\bra{\Phi}.
$$
Because the right hand side is a pure state, it must be true that
\begin{multline} \label{relocation_perBranch}
 \Tr_A [(\ket{0}_A\bra{k}A^{\R{n}} \otimes B^{\R{n}}) (C_u \ket{\Phi}_{Aa}\bra{\Phi} \otimes \ket{\psi_B}\bra{\psi_B} C_u^\dag) (\ket{0}_A\bra{k}A^{\R{n}} \otimes B^{\R{n}})^\dag]\\
  = p^{\R{n},k,\psi_B} \ket{\Phi}_{aB}\bra{\Phi}
\end{multline}
for all $\R{n}$ and $k = 0,1$, where $p^{\R{n},k,\psi_B}$ is a positive coefficient normalized by
$$
 \sum_{\R{n},k} p^{\R{n},k,\psi_B} = 1.
$$

Since Eq.\ (\ref{relocation_perBranch}) holds for any $\ket{\psi_B} \in \Hilb{B}$, we can replace $\ket{\psi_B}$ by a completely mixed state $\mathbb{I}/2$, and obtain 
\begin{multline} \label{relocation_onI}
 \Tr_A [(\ket{0}_A\bra{k}A^{\R{n}} \otimes B^{\R{n}}) (C_u \cdot \ket{\Phi}_{Aa}\bra{\Phi} \otimes \frac{1}{2}\I \cdot C_u^\dag) (\ket{0}_A\bra{k}A^{\R{n}} \otimes B^{\R{n}})^\dag]\\
 = \frac{1}{2} \Tr_A [(\ket{0}_A\bra{k}A^{\R{n}} \otimes B^{\R{n}}) (C_u  \ket{\Phi}_{Aa}\bra{\Phi} \otimes \ket{0}\bra{0} C_u^\dag) (\ket{0}_A\bra{k}A^{\R{n}} \otimes B^{\R{n}})^\dag]\\
\qquad + \frac{1}{2} \Tr_A [(\ket{0}_A\bra{k}A^{\R{n}} \otimes B^{\R{n}}) (C_u \ket{\Phi}_{Aa}\bra{\Phi} \otimes \ket{1}\bra{1} C_u^\dag) (\ket{0}_A\bra{k}A^{\R{n}} \otimes B^{\R{n}})^\dag]\\
 = (p^{\R{n},k,0} + p^{\R{n},k,1}) \ket{\Phi}_{aB}\bra{\Phi}.
\end{multline}
Note that $\ket{0}_A\bra{k}A^{\R{n}}$ acts only on Alice's input qubit.  Taking the partial trace over Alice's \textit{ancilla} qubit $\Tr_a$, Eq.\ (\ref{relocation_onI}) gives
$$
 \Tr_A [(\ket{0}_A\bra{k}A^{\R{n}} \otimes B^{\R{n}}) (C_u \cdot \frac{1}{2} \I \otimes \frac{1}{2}\I \cdot C_u^\dag) (\ket{0}_A\bra{k}A^{\R{n}} \otimes B^{\R{n}})^\dag] = \frac{(p^{\R{n},k,0} + p^{\R{n},k,1})}{2} \I,
$$
where we have used the relation
$
  \Tr_a \ket{\Phi}_{aA}\bra{\Phi} = \Tr_a \ket{\Phi}_{aB}\bra{\Phi} = \I / 2.
$
Noting that the identity operator commutes with any unitary operators, after performing the partial trace $\Tr_A$, we have
$$
 {}_A\bra{k} A^{\R{n}}A^{\R{n}\dag} \ket{k}_A B^{\R{n}}B^{\R{n}\dag} =  \frac{(p^{\R{n},k,0} + p^{\R{n},k,1})}{2} \I.
$$
This equation guarantees that Bob's accumulated operator $B^{\R{n}}$ for each sequence of measurement outcomes $\R{n}$ is proportional to a unitary operator, \ie,
\begin{equation} \label{BisUnitary}
 B^{\R{n}} = c^{\R{n}} u^{\R{n}},
\end{equation}
where the coefficient $c^{\R{n}}$ is set to satisfy
$$
 (c^{\R{n}})^2 = \frac{(p^{\R{n},k,0} + p^{\R{n},k,1})}{2{}_A\bra{k} A^{\R{n}}A^{\R{n}\dag}\ket{k}_A}.
$$

For any linear operator $T$ on $\Hilb{A}$ and the maximally entangled state given by $\ket{\Phi}_{Aa}$, $\left( T \otimes I \right )\ket{\Phi}_{Aa} = \left ( I \otimes \tr{\: T} \right ) \ket{\Phi}_{Aa}$, where $\tr{\: T}$ denotes the transpose of $T$ in the computational basis, holds.  Let $\{ S^{(i)}_x \}$ denote a set of operators forming a basis for the operators on $\Hilb{x}$ (where $x = a,A,~\mathrm{or}~B$).  That is, for any $T$ on $\Hilb{x}$, there exists a set of complex numbers $c^{(i)}$ such that
$
 T = \sum_i c^{(i)} S_x^{(i)}.
$
(An example of such a basis is the set of Pauli operators and the identity operator, if the Hilbert space in question has the dimension of 2.)  With this basis, $C_u$ on $\Hilb{A}\otimes\Hilb{B}$ can be expressed as a linear combination of $S_{A}^{(i)} \otimes S_{B}^{(j)}$, namely,
$$
 C_u = \sum_{i,j} u_{ij} S_{A}^{(i)} \otimes S_{B}^{(j)},
$$
where $u_{ij}$ denotes the coefficient of $S_{A}^{(i)} \otimes S_{B}^{(j)}$.  Let us choose $S_a^{(i)}$ to satisfy
$$
 S_a^{(i)} = \tr{S}_A^{(i)}
$$
in the computational basis and define $\tilde{C}_u$ on $\Hilb{a} \otimes \Hilb{B}$ by
$$
 \tilde{C}_u = \sum_{i,j} u_{ij} S_{a}^{(i)} \otimes S_{B}^{(j)}.
$$

Under these conventions, we have
\begin{eqnarray*}
(\ket{0}_A\bra{k}A^{\R{n}} \otimes B^{\R{n}}) (C_u \ket{\Phi}_{Aa}\bra{\Phi} \otimes \ket{\psi_B}\bra{\psi_B} C_u^\dag) (\ket{0}_A\bra{k}A^{\R{n}} \otimes B^{\R{n}})^\dag \\
=
\ket{0}_A\bra{0} \otimes \tilde{C}_u \tr{A}^{\R{n}}\ket{k}_{a}\bra{k}_a A^{\R{n}*} \otimes B^{\R{n}}\ket{\psi_B}\bra{\psi_B}B^{\R{n}\dag} \tilde{C}_u^\dag.
\end{eqnarray*}
Comparing this equation to Eq.\ (\ref{relocation_perBranch}), it must be that
\begin{equation} \label{relocation_vectorform}
 \tilde{C}_u \tr{A}^{\R{n}}\ket{k}_{a}\bra{k}_a A^{\R{n}*} \otimes B^{\R{n}}\ket{\psi_B}\bra{\psi_B}B^{\R{n}\dag} \tilde{C}_u^\dag = p^{\R{n},k,\psi_B} \ket{\Phi}_{aB}\bra{\Phi},
\end{equation}
which is equivalent to
$$
 \tilde{C}_u \tr{A}^{\R{n}}\ket{k}_{a} \otimes B^{\R{n}}\ket{\psi_B} = \exp(i \theta_{\R{n},k,\psi_B}) \sqrt{p^{\R{n},k,\psi_B}} \ket{\Phi}_{aB}.
$$
Let an ancilla state (not necessarily normalized) $\ket{v^{\R{n},k}}$ be defined by
$$
 \ket{v^{\R{n},k}} = \tr{A}^{\R{n}}\ket{k}_{a}.
$$
By substituting Eq.\ (\ref{BisUnitary}) into Eq.\ (\ref{relocation_vectorform}), we conclude that
$$
 \tilde{C}_u \cdot (\I \otimes u^{\R{n}}) \ket{v^{\R{n},k}} \otimes \ket{\psi_B} =  \exp(i \theta_{\R{n},k,\psi_B}) \sqrt{p^{\R{n},k,\psi_B}}/c^{\R{n}} \ket{\Phi}_{aB}
$$
holds for all $\ket{\psi_B}$.  The right hand side is collinear to $\ket{\Phi}_{aB}$ for all $\ket{\psi_B}$.  On the other hand, because $\tilde{C}_u \cdot (\I \otimes u^{\R{n}})$ is invertible, the left hand side returns linearly independent vectors when $\{ \ket{\psi_B} \}$ are chosen linearly independently.  This, however, is a contradiction proving that the assumption that LOCC one-piece relocalization is possible for two pieces of quantum information delocalized by the controlled-unitary operations $C_u$ must not hold.

This proof strongly depends on the fact that Bob's input state is kept arbitrary, namely, we considered the situation of delocalized two pieces of quantum information.  Indeed, if we are allowed to choose Bob's input state, one-piece relocation is possible for certain controlled-unitary operations.  An example is the controlled-NOT operation on two qubits.

\section{Conclusion and discussion} \label{conclusion}

In this paper, we first introduced the concept of pieces of quantum information and reviewed three different characterizations of the globalness of bipartite unitary operations, which were delocalization power, entanglement cost, and entangling power.  The first two characterizations are on the globalness of the unitary operations on two pieces of quantum information represented by unknown states, and the last one is on the globalness of the unitary operations on a set of known states.  We showed the fundamental difference between these two types of globalness of the unitary operations. 

Next, we extended our analysis on characterization in terms of the delocalization power by introducing a new LOCC task, LOCC one-piece relocalization of {\it one piece} of quantum information delocalized by a unitary operation.  We showed that there are unitary operations which belong to a higher globalness class in terms of the delocalization power than the local unitary equivalents of controlled-unitary operations, and such operations can be further divided into two subclasses depending on the possibility of this task. 

We also introduced another new task,  LOCC one-piece relocation of two pieces of delocalized quantum information.  We proved that LOCC one-piece relocation is impossible for any controlled-unitary operations.  This confirms that the local unitary equivalents of controlled-unitary operations, which are LOCC one-piece relocalizable, belong to a class of global operations with relatively weak globalness also in terms of LOCC relocation of quantum information.

In our analysis of the LOCC tasks, we focused on the LOCC tasks that transform two pieces of quantum information within the two-qudit Hilbert space.  This is because our main purpose is to investigate the delocalization power of two-qudit unitary operations.  But in general, we can investigate more general properties of delocalized pieces of quantum information by considering LOCC tasks that transforms $n$ pieces of quantum information delocalized in an $m$-qudit subspace of a totally $M$-qudit Hilbert space ($n \leq m$ and  $m \leq M$) to $n'$ pieces of quantum information delocalized in an $m'$-qudit subspace ($n' \le n$ and $n' \leq m'$), by limiting the allowed operations to be LOCC for a certain devision of the total Hilbert space.

Self-teleportation \cite{SelfTeleportation} can be interpreted as a special case of this generalized LOCC task for $n=n'=2$, $m=m'=2$ and $M=3$.  By denoting the total Hilbert space by $\Hilb{A} \otimes \Hilb{B} \otimes \Hilb{C}$, only LOCC is allowed between the division of $\Hilb{A}$ and $\Hilb{B} \otimes \Hilb{C}$ in this case.  It is shown that asymptotically, $k$-copies of any delocalized two pieces of quantum information in $\Hilb{A} \otimes \Hilb{B}$ can be approximately `relocated' to $\Hilb{B} \otimes \Hilb{C}$.  The error probability of this relocation drops exponentially with the number of copies $k$ as long as the two pieces of quantum information are delocalized, namely, not in a product state. 

In our analysis of the delocalization power, we characterized the order of delocalization power of unitary operations by their ability allowing the LOCC tasks.   For more quantitative analysis of the unitary operations that do not allow the LOCC tasks, it is important to analyze the entanglement cost of the corresponding entanglement assisted versions of the LOCC relocalization/relocation tasks.   

Quantum state merging \cite{StateMerging1, StateMerging2} can be interpreted as evaluating the entanglement cost for performing entanglement assisted approximate generalized LOCC task for $n=n'=3$, $m=m'=3$ and $M=4$ ($\Hilb{A} \otimes \Hilb{B} \otimes \Hilb{C} \otimes \Hilb{R}$), where only LOCC is allowed between the division of $\Hilb{A}$ and $\Hilb{B} \otimes \Hilb{C}$ and no operation is allowed on $\Hilb{R}$.   This is an entanglement assisted LOCC task to achieve relocation of three pieces of quantum information delocalized in $\Hilb{A} \otimes \Hilb{B} \otimes \Hilb{R}$ to $\Hilb{B} \otimes \Hilb{C} \otimes \Hilb{R}$.  In the asymptotic limit, it is shown that the entanglement cost coincides with the quantum conditional entropy, which provides an operational interpretation of the quantum conditional entropy.   To understand information theoretical meanings of our LOCC tasks, it is interesting to analyze asymptotic settings of our LOCC and entanglement assisted LOCC tasks.   We leave these investigations for future works.  

\begin{acknowledgments}
This work is supported by the Special Coordination Funds for Promoting Science and Technology, Institute for Nano Quantum Information
Electronics, and by Global COE Program ``the Physical
Sciences Frontier'', MEXT, Japan.
\end{acknowledgments}

\end{document}